\documentstyle[pre,aps,epsf,floats,twocolumn]{revtex}

\begin{document}
\draft
\title{Dissipative chaotic scattering}
\author{Adilson E. Motter$^{1,*}$ and Ying-Cheng Lai$^{1,2}$}
\address{1. Department of Mathematics, Center for Systems Science
and Engineering Research, Arizona State University, Tempe, Arizona 85287}
\address{2. Departments of Electrical Engineering and Physics, 
Arizona State University, Tempe, Arizona 85287}
\date{Phys. Rev. E {\bf 65}(1) (2002), 015205 (Rapid Communication)}
\maketitle

\begin{abstract}
We show that weak dissipation, typical in realistic 
situations, can have a metamorphic consequence on nonhyperbolic chaotic
scattering in the sense that the physically important particle-decay law 
is altered, no matter how small the amount of dissipation. As a result,
the previous conclusion about the unity of the fractal dimension of the set
of singularities in scattering functions, a major claim about nonhyperbolic
chaotic scattering, may not be observable.
\end{abstract}


Chaotic scattering \cite{Early_papers,Chaos_focus,Gaspard:book} is a physical manifestation
of transient chaos \cite{GOY:1983}, which is due to the existence of 
nonattracting chaotic invariant sets, i.e., chaotic saddles, in the phase space.
As a result, a Cantor set of singularities arises in physically measurable
scattering functions relating an output variable after the scattering to an
input variable before the scattering \cite{Chaos_focus}. 
Generally, the dynamics of chaotic scattering may be characterized as 
either {\it hyperbolic} or {\it nonhyperbolic}. In hyperbolic chaotic scattering, 
all the periodic orbits are unstable and there are no Kol'mogorov-Arnol'd-Moser (KAM)
tori in the phase space and, as such, the survival probability of a particle in the
scattering region typically decays exponentially with time.
In nonhyperbolic chaotic scattering \cite{Nonhyp_scatt,LFO:1991}, there are both 
KAM tori and chaotic sets in the phase space. Due to the stickiness effect
of KAM tori, a particle initialized in the chaotic region can spend a long time 
in the vicinity of KAM tori, leading to an algebraic decay \cite{Algebr_decay} 
of the survival probability of the particle in the scattering region. 
A surprising result in nonhyperbolic chaotic scattering is that, because of the 
algebraic decay, the fractal dimension of the set of singularities in a scattering
function is unity \cite{LFO:1991}.  

A physically important issue in the study of nonlinear dynamics is 
to understand how robust a phenomenon is against perturbations or deviations
between the underlying mathematical model and physical reality.  
In the case of chaotic scattering, most of the theoretical investigations 
so far have been restricted to Hamiltonian or conservative systems. 
In a realistic situation, a small amount of dissipation can be expected.
Take, for example, chaotic scattering arising in the context of particle
advection in hydrodynamical flows \cite{Fluid_dynamic}. In most  existing
studies, the condition of incompressibility is assumed of the underlying flow
\cite{Fluid_dynamic}, which allows the problem to be casted in the  context of
Hamiltonian dynamics as the particle velocities can be related to flow's
stream function in a way that is completely analogous to the  Hamilton's
equations in classical mechanics. Real hydrodynamical flows cannot be
perfectly incompressible, and the effects of inertia and finite mass of  the
particles advected by the flow are effectively those due to friction,  or
dissipation \cite{MR:1983}. 

The aim of this paper is to study the effect of dissipation on chaotic scattering
dynamics. We first consider hyperbolic chaotic scattering and argue that
weak dissipations have a negligible effect on the physical observables of
chaotic scattering, such as scattering functions. We then focus on 
nonhyperbolic chaotic scattering and find that, in contrast to the hyperbolic
case, the scattering dynamics can be altered by weak dissipation in a
fundamental way. The major consequence of dissipation is  that it typically
converts KAM tori into periodic attractors. As a result, the
underlying chaotic saddle can undergo a metamorphic bifurcation to a
structurally different chaotic set, playing the role of chaotic invariant set
that generates fractal basin boundaries \cite{GMOY:1983}. There is an
immediate transformation of the decay law of scattering particle from being
algebraic in the Hamiltonian case to being exponential in the dissipative 
case, no matter how small the amount of dissipation. As a result, the
fractal dimension of the chaotic saddle decreases from the integer value in
the Hamiltonian case. These findings have striking
implications to the study of chaotic scattering: they suggest that the 
algebraic-decay law, regarded to hold universally in nonhyperbolic chaotic
scattering, is apparently structurally unstable against weak dissipations. 
More importantly, the previously believed integer dimensions of the chaotic 
saddles \cite{LFO:1991} in nonhyperbolic chaotic scattering may not be
observable in realistic physical situations where dissipation is present.

We begin by presenting a picture for the formation of the fractal
sets in chaotic scattering. Consider the idealized model of the hierarchical
construction of Cantor sets in the unit interval. For hyperbolic scattering,  
an open subinterval in the middle of the unit interval is removed first.
From each one of the two remaining subintervals, the same fraction from 
their middle is removed, and so on. Each step of this construction
can be thought of as an iteration of the hyperbolic tent map with slope
larger then two. The total length that remains decays exponentially with the
number of iterations and the resulting Cantor set has a fractal dimension
(e.g., box-counting dimension) smaller than one. For nonhyperbolic chaotic
scattering, the same construction applies but the fraction removed at each
step decreases with time, say, is inversely proportional to time. This simple
reduction of the fraction removed captures the essence of the effect of KAM
tori: their ``stickiness'' to particle trajectories in the phase space
\cite{LFO:1991}. The remaining length decays algebraically with time and, even
though the measure of the remaining set asymptotes to zero, the resulting
Cantor set has dimension one \cite{LFO:1991}.

How does dissipation change the above construction? The skeleton of the
underlying chaotic saddle is formed by periodic orbits. When the system is
hyperbolic, its structural stability guarantees the survival of all periodic
orbits under small changes of the system parameters.
Accordingly, the structure of the Cantor set in the presence of a small amount
of dissipation is expected to be the same as before. When the dynamics is
nonhyperbolic, however, qualitatively different behavior can take place.
Marginally stable periodic orbits in KAM islands can become stable, turning
their nearby phase-space regions into the corresponding basins of attraction
\cite{FG:1997}. This means that, part of the previous chaotic saddle now
becomes part of the basins of the attractors. Most importantly for the
scattering  dynamics, the converted subset supports orbits in the
neighborhood of the KAM islands that otherwise would be scattered after a
long, algebraic time. These orbits are solely responsible for the nonhyperbolic
character of the scattering in the conservative case. Due to the existence of
dense orbits in the original chaotic saddle, the noncaptured part of the
invariant set remains in the boundaries of basins of the periodic attractors.
Therefore, the invariant set is the asymptotic limit of the boundaries
between scattered and {\it captured} orbits, rather than those between
scattered and {\it scattered} orbits as in the conservative case. Chaos thus
occurs on the nonattracting invariant set whose stable manifold becomes the
boundary separating the basins of the attractors and of the scattering
trajectories. Through this simple reasoning we can see that the structure and
the meaning of the Cantor set is fundamentally altered: in successive steps we
remove a {\it constant} instead of a decreasing fraction in the middle of each
interval. As a result, the scattering dynamics becomes hyperbolic with
exponential decay.  The dimension of the Cantor set immediately decreases from
unity  as a dissipation parameter is turned on. There is now more than one
possible outcome: some of the removed intervals correspond to scattered
orbits and the others correspond to orbits captured by the attractors. We
stress that the appearance of attractors accompanied by a metamorphosis of the
chaotic saddle can occur for arbitrarily small dissipation.

We now present numerical support for the effect of weak dissipation on 
chaotic scattering, particularly the metamorphic transformation of particle
decay and fractal dimension in the nonhyperbolic case. Our model is a 
dissipative version of the two-dimensional area-preserving map utilized in
Ref. \cite{LFO:1991} to establish the unity of the fractal dimension, 
a particularly convenient model for studying nonhyperbolic chaotic scattering.
The map reads
\begin{equation} \label{eq:map}
M\left( \begin{array}{ll}
x \\ y \end{array} \right) 
= \left\{ \begin{array}{ll}
\lambda [x-(x+y)^2/4-\nu (x+y)]\\
\lambda^{-1}[y+(x+y)^2/4]
\end{array} \right.,
\end{equation}
where $\lambda>1$ and $\nu\ge 0$ are parameters.
The map is conservative for $\nu=0$ and dissipative for $\nu>0$.
For $\nu=0$, almost all orbits started from negative values of $y$ are
scattered to infinity. In this case, the dynamics is nonhyperbolic for
$\lambda\lesssim 6.5$ and hyperbolic for $\lambda\gtrsim 6.5$. The
computation of the particle decay and fractal dimension in nonhyperbolic
scattering requires examining very small scales, which makes the numerical
computation a highly nontrivial task \cite{Nonhyp_scatt}. The advantage of
using map (\ref{eq:map}) instead of a continuous flow is that it makes
high-precision computation possible. Our results are, however, 
expected to hold in typical nonhyperbolic systems with KAM tori.

\begin{figure}[h]
\begin{center}
\epsfxsize=10pc
\epsfbox{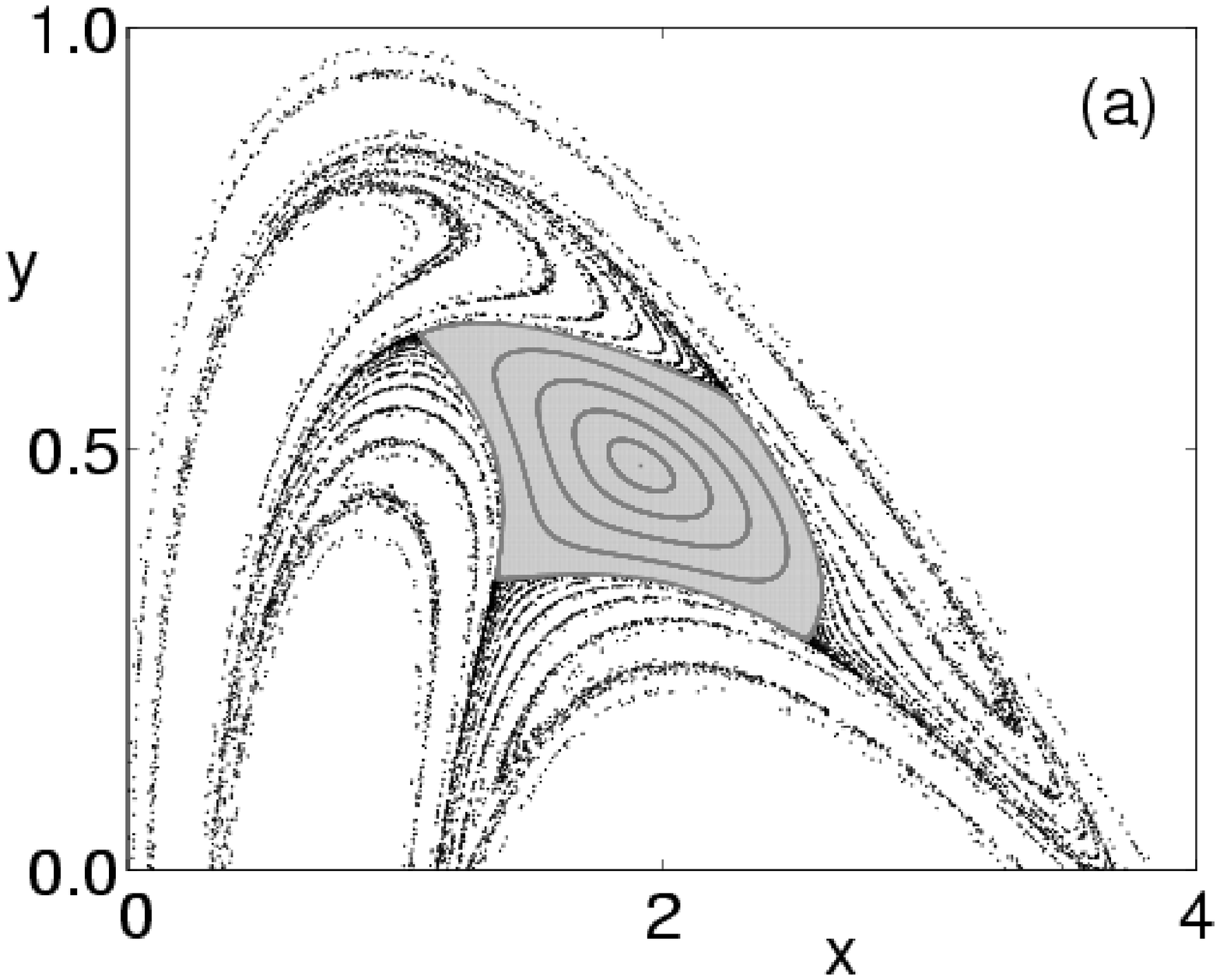}
\epsfxsize=10pc
\epsfbox{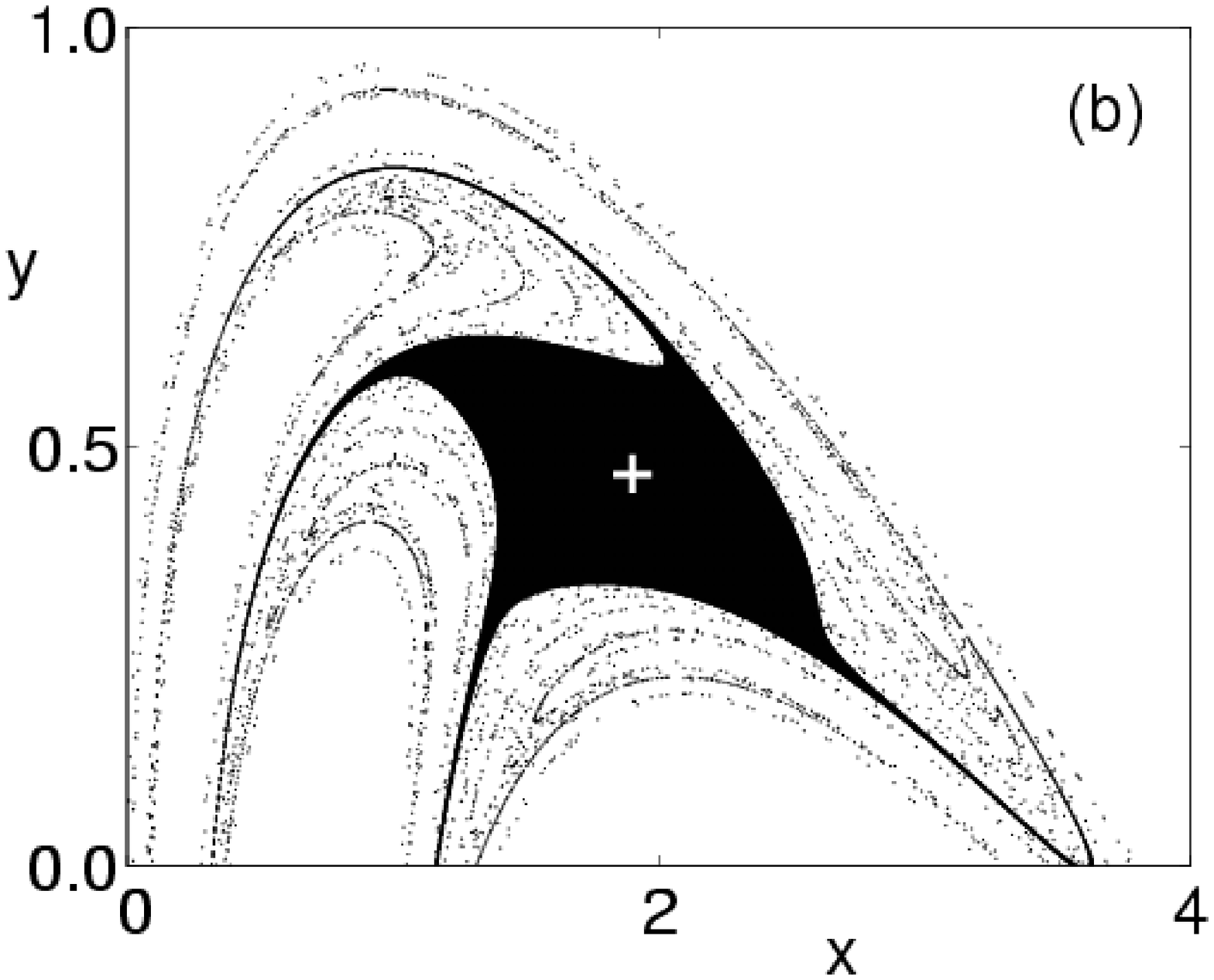}
\epsfxsize=10pc
\epsfbox{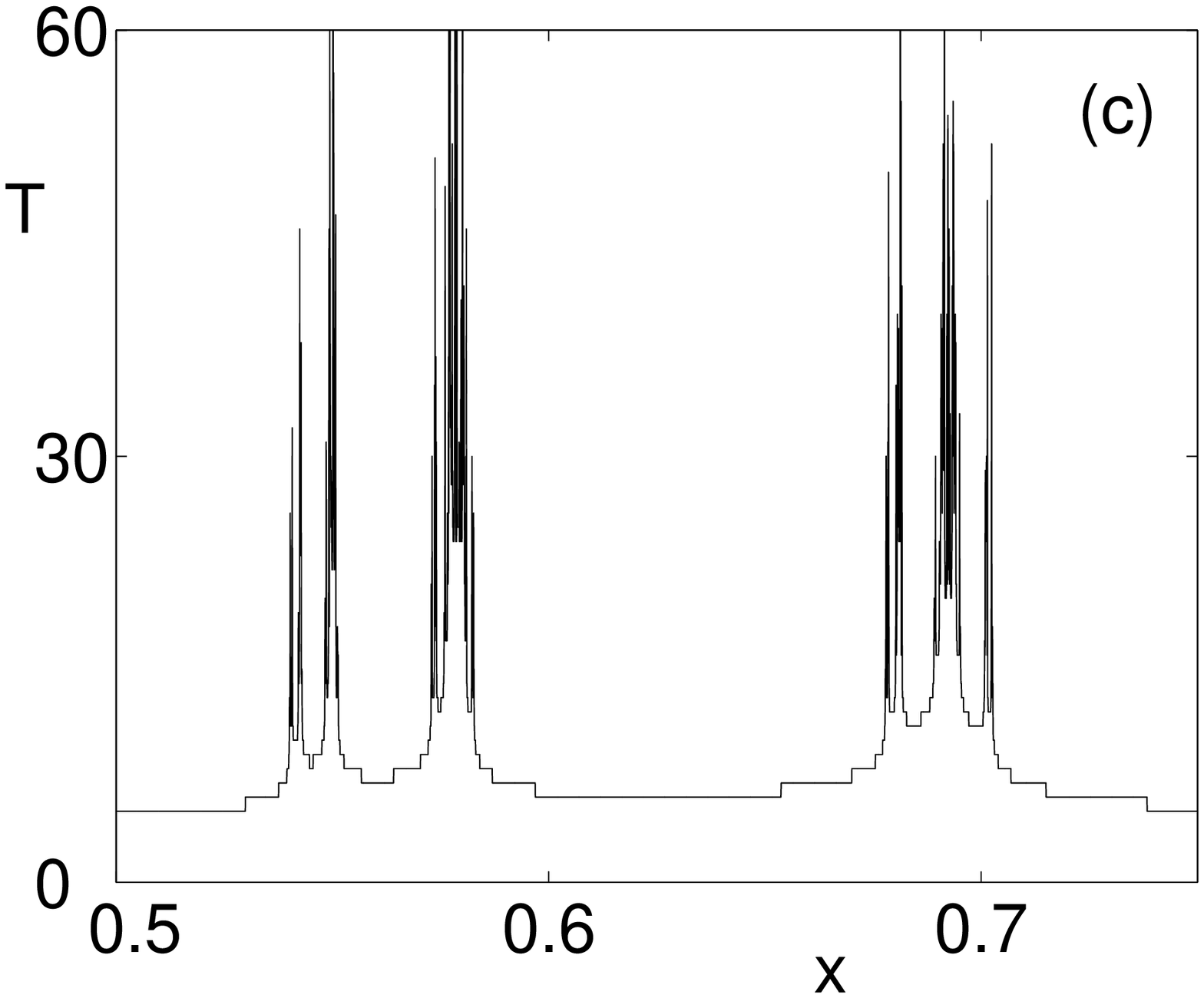}
\epsfxsize=10pc
\epsfbox{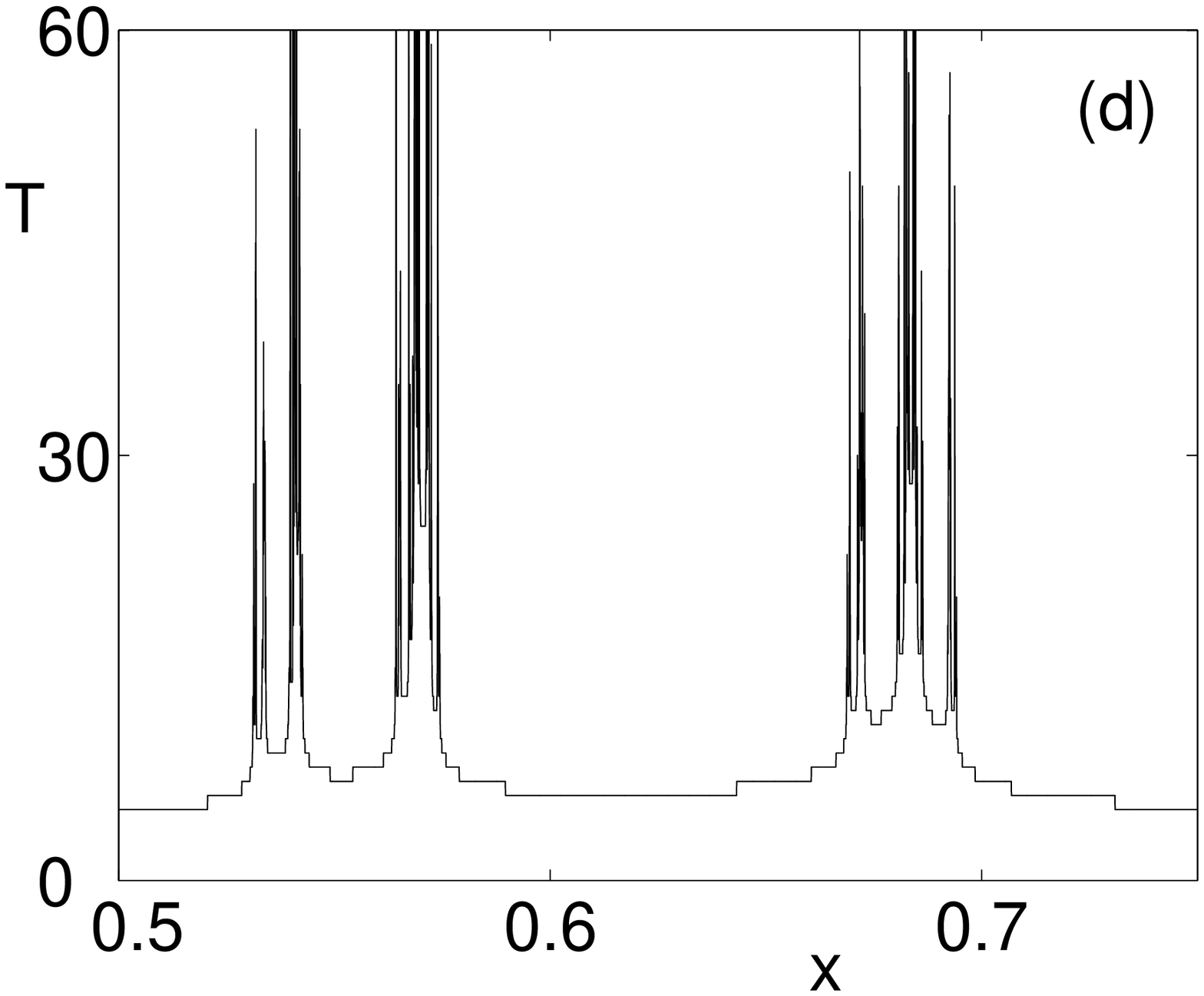}
\vspace{0.5cm}
\caption{Phase-space structure and time-delay function for map (\ref{eq:map})
with $\lambda=4.0$. (a) $\nu=0$: KAM island (in grey), scattered
orbits (in blank), and the fractal boundaries of the scattered orbits (in
black). (b) $\nu=0.01$: captured orbits (in black) and scattered
orbits (in blank). The plus sign is the fixed point attractor.
(c),(d) Time delay in the conservative and dissipative cases of
(a) and (b), respectively.
$T$ is the time taken by particles to reach $\sqrt{x^2+y^2}\geq 100$.
}
\label{fig:nonhyperbolic}
\end{center}
\end{figure}

We study map (\ref{eq:map}) in the nonhyperbolic regime,
without and with dissipation. We set
$\lambda=4.0$. When there is no dissipation ($\nu=0$), there is a major KAM
island in the phase space, as shown in Fig. \ref{fig:nonhyperbolic}(a). The
fractal boundaries of the basins of scattering trajectories to infinity 
are also shown, which correspond to the stable manifold of the chaotic saddle
in the scattering region. When dissipation is present ($\nu>0$), the fixed point  
in the center of the island becomes an attractor. Dynamically, it happens
because the magnitudes of the eigenvalues of periodic orbits associated 
with islands are one, which are reduced by dissipation in general. 
The basin of attraction of this attractor ``captures''
the island itself and orbits close to the stable manifold
of the previously existing invariant set, as shown in Fig. \ref{fig:nonhyperbolic}(b). 
The intricate character of the basin of attraction with apparent fractal
boundaries comes from points of the invariant set that are arbitrarily close
to the island for $\nu=0$. The newly created basin of attraction contains
these points and hence, all their preimages as well. These preimages extend
in the phase space along the original  stable manifold of the chaotic saddle,
which is the reason that the boundaries mimic those of the original basins of
scattering trajectories [Figs. \ref{fig:nonhyperbolic}(a) versus
\ref{fig:nonhyperbolic}(b)]. Because of this similarity, the scattering
functions and time-delay functions, which are physically measurable, {\it
resemble} each other in both the conservative and weakly dissipative case, as
shown, respectively, in Figs. \ref{fig:nonhyperbolic}(c) and 
\ref{fig:nonhyperbolic}(d), where the time delay of particles launched from
the horizontal line $y = -2$ toward the scattering region is plotted against
their $x$-coordinates on the line.
    
\begin{figure}[h]
\begin{center}
\epsfxsize=10pc
\epsfbox{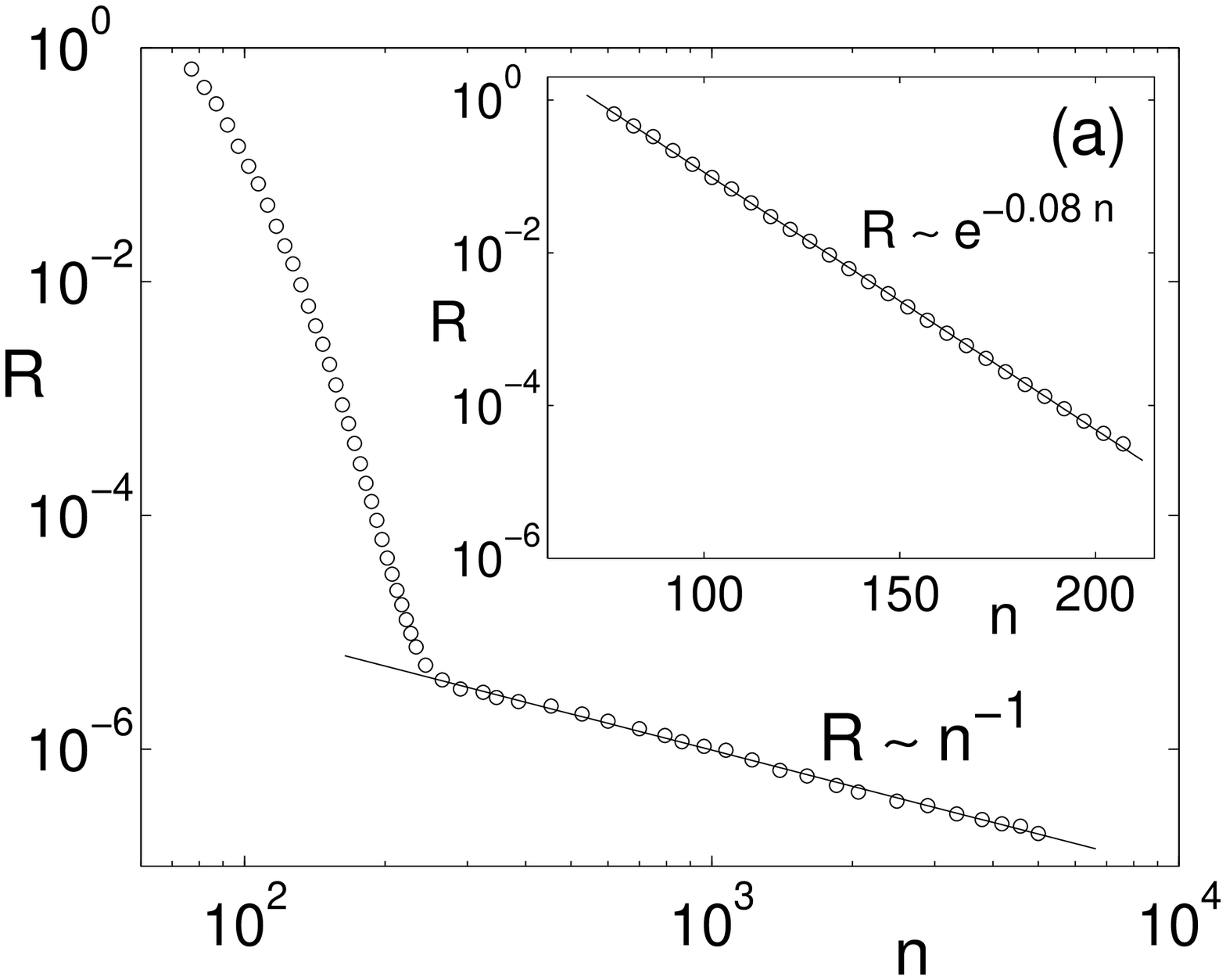}
\epsfxsize=10pc
\epsfbox{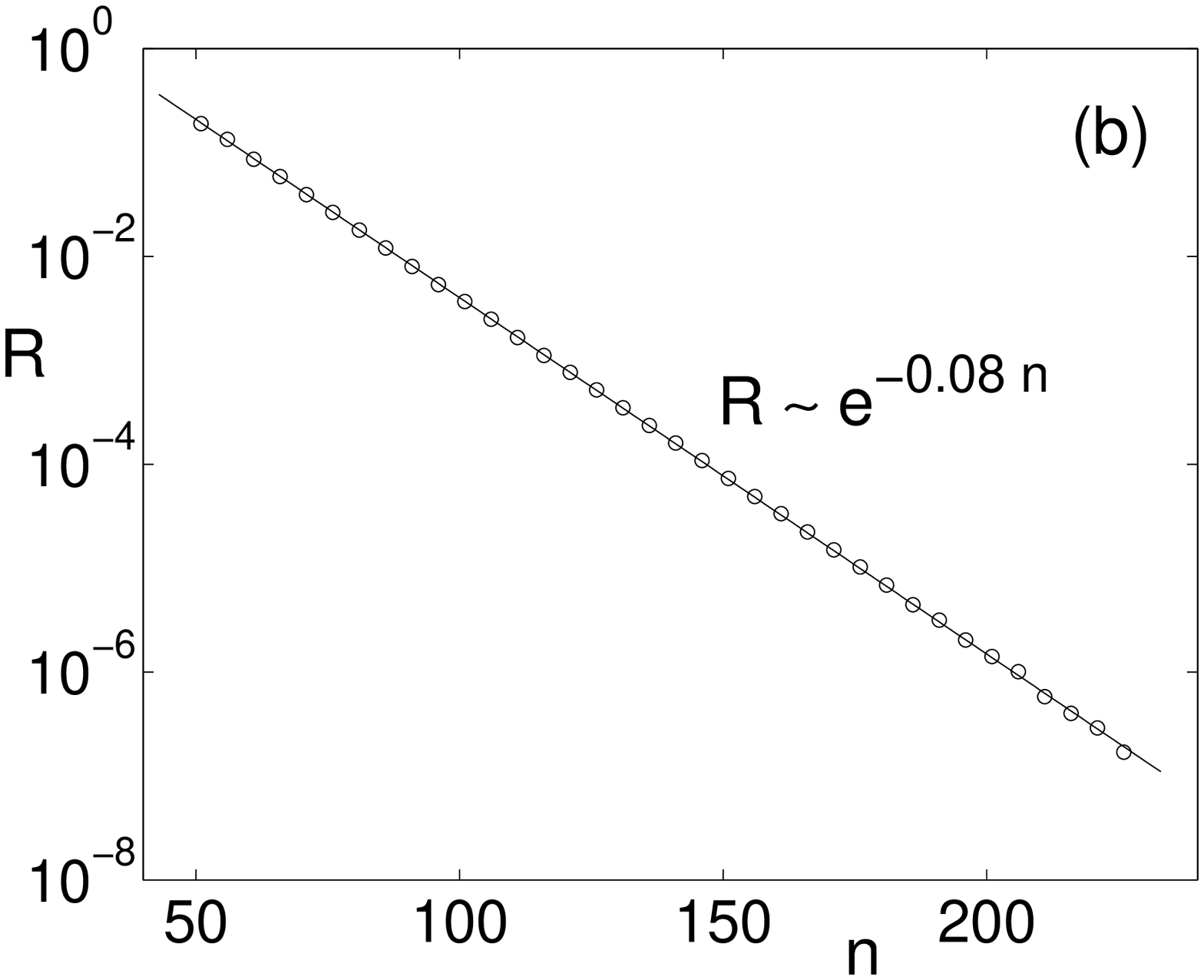}
\vspace{0.5cm}
\caption{Time decay for map (\ref{eq:map}) with $\lambda=4.0$,
in the interval $[x_0,x_0+10^{-7}]$. (a) $\nu=0$, $x_0=0.5770050$.
The inlet corresponds to the initially exponential decay.
(b) $\nu=0.001$, $x_0=0.5760006$.}
\label{fig:decay}
\end{center}
\end{figure}

To examine the decay laws of the scattering particles, we approximate the
survival probability of a  particle in the scattering region by $R(n)$, the
fraction of a large number of particles still remaining in the scattering
region  (defined by $\sqrt{x^2+y^2}<r$) at time $n$, which are  
initiated in subregions close to the boundaries of the scattering basins.
For convenience, we choose $r=100$ and choose
initial conditions from the horizontal line at $y_0=-2$. When the dynamics is
nonhyperbolic and  conservative, the decay of $R(n)$ with time is exponential
for small $n$ and algebraic for large $n$, as shown in Fig. \ref{fig:decay}(a)
for $\lambda=4.0$: $R\sim e^{-\alpha n}$ for $n\lesssim 250$ and $R\sim
n^{-\beta}$ for $n\gtrsim 250$, where $\alpha\cong 0.08$ and $\beta\cong 1.0$.
In the presence of a small dissipation, the time decay becomes strictly exponential
and with the same decay rate of the exponential regime of the conservative
case, $\alpha\cong 0.08$, as shown in Fig. \ref{fig:decay}(b) for $\nu=0.001$. 
The original algebraic decay in the conservative case is destroyed by  
the dissipation because orbits with points close to the island,
and that otherwise would be stuck, are captured by the periodic attractor. 
The decay rate in general
changes under further increases of the dissipation. For $\nu=0.01$, for
instance, we obtain $\alpha\cong 0.06$. In the hyperbolic region the time
decay is always exponential. For $\lambda=8.0$, for example, the decay rate
$\alpha$ remains essentially constant and equal to $0.9$ in the range
$0\leq\nu\leq 0.01$.

The uncertainty algorithm \cite{GMOY:1983} can be used to compute the fractal
dimension $D$ of the set of intersection points between the stable manifold
of the chaotic saddle and a line from which scattering particles are initiated. 
As above, we choose the line $y_0=-2$. In
the absence of attractors, $D$ is the dimension of the set of singularities in
scattering functions. In Ref. \cite{LFO:1991}, it is argued that $D=1$ when
map (\ref{eq:map}) is nonhyperbolic and conservative. A technical point about
the numerical evaluation of the dimension in this case is that the result
converges {\it slowly} to unity, and the convergence rate is determined by
the reduction of length scales in the computation \cite{LFO:1991}. 
When a small amount of dissipation
is present, $D+1$ becomes the dimension of the boundaries between scattered
and captured basins. The numerical convergence of $D$ is in this case {\it faster}
and essentially independent of the size of the interval under consideration.
For $\lambda =4.0$ and $\nu=0.01$, we obtain $D \cong 0.8$, a well-convergent
value as the length scale is reduced over six orders of magnitude.
The dimension is much less sensitive to the presence of dissipation if the
dynamics is hyperbolic. For $\lambda=8.0$, for instance, we obtain $D\cong
0.44$ for both $\nu=0$ and $0.01$.

Dissipation can also lead to several coexisting attractors for some values of
$\lambda$ in the originally nonhyperbolic region. Periodic
attractors are created through saddle-node bifurcations as $\lambda$ is
varied. These attractors then undergo period doubling cascades until the
accumulation point where chaos appears. For small dissipation, however, the
chaotic interval in the parameter space can be so small that it is difficult to
detect chaotic attractors numerically. In fact, for $\nu$ on the order of $0.01$ or
smaller, the dynamics of map (\ref{eq:map}) is dominated by low-period periodic attractors.
Periodic attractors of high periods either have small basins of attraction or
exist in small intervals in the parameter space. Since periodic attractors result
from the stabilization of periodic orbits in KAM islands, the parameter regions in 
which these attractors exist are approximately the same as those of the 
corresponding islands.
In addition, the sizes of the basins are of the same order of the sizes of the
original islands in the phase space \cite{noise}.

In summary, our qualitative and quantitative examinations indicate that 
weak dissipation, no matter how small, can fundamentally alter the nature
and dynamics of nonhyperbolic chaotic scattering. The algebraic-decay law,
commonly believed to hold in such a case, is typically converted into 
an exponential-decay law in the metamorphic sense that the conversion 
can be induced by arbitrarily small amount of dissipation. A consequence of
such a metamorphosis is that the previously claimed \cite{LFO:1991}  unity of
the fractal dimension of the set of singularities in scattering functions may
not be physically meaningful. To our knowledge, there has been no previous
attempt to address the effect of dissipation in open Hamiltonian systems, but
this is a physically important issue of nonlinear dynamics that deserves
further attention. 

This work was supported by AFOSR under Grant No. F49620-98-1-0400. 
AEM acknowledges financial support from FAPESP.

\end{document}